\documentclass[fleqn,twoside]{article}
\usepackage{graphicx,espcrc2}

\title{Muon spin rotation study of the $(TMTSF)_2 ClO_4$ system}

\author{A.J. Greer\address[GU]{Physics Department, Gonzaga University, Spokane, WA 99258-0051, USA},
D.R. Harshman\address[PHY]{Physikon Research Corporation, PO Box 1014, Lynden, WA 98264-1014, USA}\address[UND]{Department of Physics, University of Notre Dame, Notre Dame, IN 46556, USA},
W.J. Kossler\address[WM]{Physics Department, College of William and Mary, Williamsburg, VA 23187-8795, USA},
A. Goonewardene\addressmark,
D.Ll. Williams\address[UBC]{Department of Physics, University of British Columbia, Vancouver, BC V6T 1Z1, Canada},
E. Koster\addressmark,
W. Kang\address[UCG]{Department of Physics, University of Chicago, Chicago, IL 60637, USA},
R.N. Kleiman\address[LT]{Lucent Technologies, Bell Labs, Murray Hill, NJ 07974, USA},
and R.C. Haddon\address[UCR]{Department of Chemistry, University of California at Riverside, Riverside, CA 92521, USA}}

\begin{document}

\begin{abstract}
We report a study of the organic compound $(TMTSF)_2 ClO_4$ in both
a sample cooled very slowly through the anion ordering temperature
({\it relaxed} state) and a sample cooled more rapidly ({\it intermediate}
state).  For the {\it relaxed} state the entire sample is observed to be
superconducting below about $T_{\rm c}\simeq 1.2\,$K.  The second moment
of the internal field distribution was measured for the relaxed state 
yielding an in-plane penetration depth of $\simeq 12000\,$\AA.  The
{\it intermediate} state sample
entered a mixed phase state, characterized by coexisting macroscopic sized
regions of superconducting and spin density wave (SDW) regions, below
$T_{\rm c}\simeq 0.87\,$K.
These data were analyzed using a back-to-back cutoff exponential function,
allowing the extraction of the first three moments of the magnetic field
distribution.  Formation of a vortex lattice is observed 
below $0.87\,$K as evidenced by the diamagnetic shift for the two fields
in which we took {\it intermediate} state data.
\end{abstract}

\maketitle

\noindent{{\it Key words:} organic superconductor, penetration depth, gap
symmetry, spin density wave

\section*{Introduction}
The organic superconductors $(TMTSF)_2 X$ (Bechgaard salts, where
$TMTSF$ means tetramethyltetraselenafulvalene), and where $X$ is $PF_6$,
$ClO_4$, {\it etc.}, have been under much scrutiny of late due to
their rich array of magnetic behavior\cite{Belain,Cooper,Jerome,Osada,Hannahs}.
The $PF_6$ compound undergoes
a metal-insulator transition at ambient pressure and below $12\,$K
enters a spin density wave (SDW) state.  Under relatively low
pressure, this material goes into a superconducting state with
$T_{\rm c} \simeq 1.1\,$K\cite{Ishiguro,Harsh1}.
The $ClO_4$ compound, however, becomes superconducting at
ambient pressure if it is cooled slowly enough through the anion ($ClO_4$)
ordering temperature, $\simeq 24\,$K\cite{Bechgaard}.  The slow cooling
through the ordering temperature allows the anions to become ordered, and
this is termed the {\it relaxed state} of the material.  If, in contrast,
the $ClO_4$ compound is cooled
rapidly through the ordering temperature it enters a SDW phase below
about $4\,$K.  Between these two extremes are a whole series of
{\it intermediate states}.  These are achieved by first slowly cooling the
sample from about $40\,$K down to a quenching temperature, and then
quenching the sample by rapidly cooling to liquid helium temperatures at a rate
of at least $60\,$K/min.  When the quenching temperature is varied around the
anion ordering temperature of $24\,$K\cite{Ishiguro}, 
a partial degree of anion order results.  Samples treated in this manner
have been observed to have coexisting, macroscopically sized regions of
both superconducting and SDW phases\cite{Schwenk}.

Recently there has been much investigation into the superconducting state
of these materials.  As with the high-$T_{\rm c}$'s, much of the interest
has focused on the nature of the pairing state of the superconducting
quasi-particles.  In the early 80's there was evidence in favor of both
conventional
BCS-like pairing\cite{Belain,Garoche,Chaikin,Murata} as well as $p$-wave
pairing\cite{Choi,Bouffard,Coulon,Tomic}.  More recently there has been much
evidence from upper critical field measurements\cite{Lee2,Lee3}, Knight shift
measurements \cite{Lee4,Lee5}, and corresponding theoretical
support\cite{Duncan,Lebed,Kuroki,Miyazaki}
suggesting that the pairing state may be triplet, or $f$-wave, in nature.

Reported here are the results of one of the few studies employing $\mu$SR
to investigate this material.  We present results of {\it relaxed} $ClO_4$ in
a transverse magnetic field of $300\,$Oe.  Also presented are results of
an {\it intermediate} state at two different transverse magnetic fields,
$100\,$Oe and $190\,$Oe.

\section*{Experiment}
All samples were made by the usual electrochemical technique\cite{Ishiguro}.
The first sample consisted of long needles of typical dimensions 
$2\,$cm $\times$ $2\,$mm $\times$ $0.5\,$mm along the {\bf a}, {\bf b}, and
{\bf c} directions, respectively, and was studied in the {\it relaxed} state.
The second was assembled from pieces of typical dimensions
$5\,$mm $\times\,1\,$mm $\times\,0.5\,$mm and was studied in the
{\it intermediate}
state.  For both samples, the crystals were oriented into a mosaic with axes
parallel.

The data were acquired at the M15 muon beam line at the TRIUMF cyclotron
facility in Vancouver, BC, Canada.  The samples were mounted on 99.999\%
pure annealed silver (in which muons do not depolarize)
and cooled in a dilution refrigerator.  
The {\it relaxed} state sample was first cooled from $80 \rightarrow 32\,$K
at a rate of $267\,$mK/min and then from $32 \rightarrow 15\,$K at a rate of
$30\,$mK/min.
The {\it intermediate} state sample was cooled at a rate of $50\,$mK/min 
from $30 \rightarrow 16\,$K through the anion ordering temperature.
Cooling was again done in zero external field.

All data were acquired with the external field applied parallel to
the crystal {\bf c} axis.  The initial muon polarization direction was
approximately perpendicular to the applied field in a standard transverse
field geometry.  For a more complete discussion of the $\mu$SR technique,
see, {\it e.g.}, \cite{Schenck}.

\section*{Results for the {\it relaxed} state sample}
The analysis of the {\it relaxed} data was performed using the following
Gaussian relaxation function:
\begin{eqnarray}
G(t)=A e^{- \sigma^2 t^2} \cos(\omega t + \phi)
\end{eqnarray}
We have omitted the $1/2$ factor in the exponent for simplicity, and we used
a single component function because the large sample size prevented any
appreciable background in the data.  The results of fits with this function
are shown in Fig. \ref{fig1}.
\begin{figure}[htb]
\includegraphics[width=7.5cm]{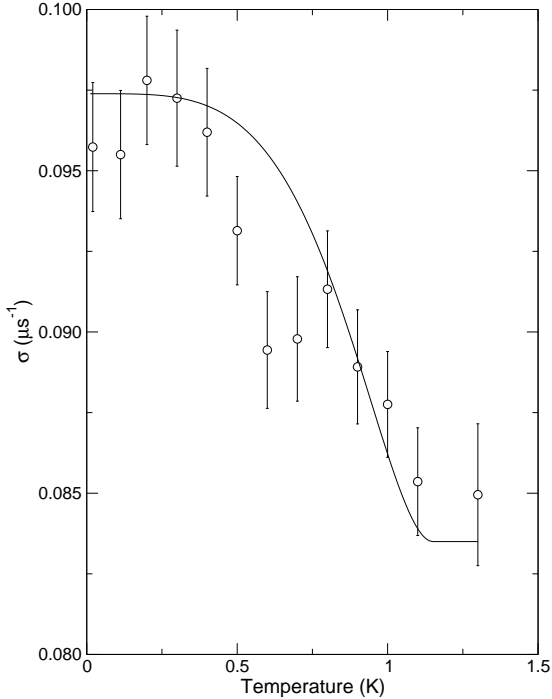}
\caption{Plot of the relaxation rate $\sigma$ as a function of temperature
for the {\it relaxed} (superconducting) state at $H_{ext}=300\,$Oe.  The
curve is
a fit using the two-fluid model assuming a baseline of $0.0835\,\mu$s$^{-1}$
and $T_{\rm c}=1.15\,$K, yielding $\lambda(T\rightarrow 0\,$K$)=12300\,$\AA.}
\label{fig1}
\end{figure}

The superconducting transition is clearly evident at about $1.1\,$K, where
the formation of a flux lattice develops.  This is consistent with previously
published values for $T_{\rm c}$\cite{Ishiguro,Harsh1}.  The increase in the
relaxation rate $\sigma$ with decreasing temperature indicates a broadening
of the field distribution due to the formation of a flux line lattice (FLL)
in the superconducting state.  The shape of this plot can often yield
information on the pairing state of the superconducting
quasiparticles\cite{Tinkham}.  The curve through the data is a guide to the
eye using the two-fluid model for the temperature
dependence of the penetration depth.  That is \cite{Brandt}:
\begin{equation}\label{secmom}
\langle (\Delta B)^2 \rangle = \frac{2 \sigma^2}{\gamma_{\mu}^2}=\frac{0.00371 \phi_o^2}{\lambda^4(T)}
\end{equation}
and with
\begin{equation}
\lambda(T)=\lambda(T=0) \left[1-\left(\frac{T}{T_{\rm {c}}}\right)^4\right]^{-1/2}
\end{equation}
Here, $\phi_o=2.068\times 10^{-7}\,$G cm$^2$ is the flux quantum,
$\langle (\Delta B^2) \rangle$ is the second moment,
and $\gamma_{\mu}=85.137\,$Mrad/s/kG is the muon gyromagnetic ratio.
For equation \ref{secmom} a triangular lattice is assumed, and the value for
$\sigma$ is found from Fig. \ref{fig1} by subtracting the above $T_{\rm c}$
value from the $T \rightarrow 0\,$K value in quadrature.  The low temperature
penetration depth for these data was previously determined to be
$\lambda_{ab}=12000\pm 2000\,$\AA \cite{Harsh1}, where $\lambda_{ab}$ is an
average penetration depth obtained from equation \ref{secmom}.
Since the penetration depths along {\bf a} and {\bf b} are different this
$\lambda_{ab}=\sqrt{\lambda_a \lambda_b}$.  Field distributions expected when
$\lambda_a \neq \lambda_b$ can be seen in ref\cite{Book}.
Our data appear to become temperature independent as $0\,$K is approached,
which is consistent with $s$-wave pairing; however, it is not possible
to precisely determine the pairing state due to the size of the error
bars.

The data here, in contrast to those of L.P. Le {\it et al.}\cite{Le}, have
many more low temperature points.  The clear rise in relaxation rate, seen in
our data below about $1\,$K, may have even been seen in their data, but was
discounted due presumably to the sparseness and lower statistics of their
data.

\section*{Results for the sample in the {\it intermediate} state}
Before taking data and lowering the sample temperature, an external field of
$100\,$Oe was applied and held fixed during this phase of the experiment.
The analysis was performed using a back-to-back exponential function of the
form\cite{Harshman}
\begin{equation}
n(\omega)=\left\{  \begin{array}{ll}
a_L e^{(\omega - \omega_p)\tau_L} & \mbox{ $(\omega < \omega_p)$ }\\
a_R e^{(\omega_p - \omega)\tau_R} & \mbox{ $(\omega > \omega_p)$ }
\end{array} \right.
\end{equation}
to represent the fields associated with superconductivity.  Here
$\omega_p$ is the frequency of the peak of the frequency
distribution, $\omega$ is the frequency, $\tau_L$ and $\tau_R$ are
the decay factors to the left and right of the peak frequency, $a_L$ and
$a_R$ are constants, and $n(\omega)$ is the probability per unit frequency
interval of a given frequency.  An assumed Gaussian-like distribution of
fields arising from nuclear dipoles was convoluted with this.  This convoluted
form has an analytical Fourier transform which was used as the first component
of a two component fitting function.  The second ({\it background}) component
is attributed to muons which do not stop in the sample.  This sample was of
a smaller size than the {\it relaxed} sample, and we found it appropriate
to include a background signal.  The overall fitting function is shown below.
\begin{eqnarray}
\lefteqn{G(t)= A_1 B(t) e^{-(\sigma_1 t)^2}+ } \hspace{.5in} \nonumber \\ 
& & A_2 e^{-(\sigma_2 t)^2}\cos(\omega_2 t+\phi) 
\end{eqnarray}
Subscripts here refer to first and second components.  Also,
\begin{eqnarray}
\lefteqn{B(t)=(r_1(t)+r_2(t)) \cos(\omega_1 t+\phi) + } \hspace{.25in}\nonumber \\
& & t(r_1(t)/\tau_L - r_2(t)/\tau_R)\sin(\omega_1 t+\phi) 
\end{eqnarray}
and
\begin{eqnarray}
r_1(t)=\frac{\tau_R}{(\tau_L+\tau_R)(1+(t/\tau_L)^2)} \nonumber \\
r_2(t)=\frac{\tau_L}{(\tau_L+\tau_R)(1+(t/\tau_R)^2)}
\end{eqnarray}
The nuclear dipole field spread parameter, $\sigma_1$, and the {\it background}
parameters ($A_2$, $\sigma_2$, and $\omega_2$) were determined
by fits above $T_{\rm c}$ and were held fixed for subsequent fits.  (For the
$190\,$Oe data these parameters were found {\it at} $T_{\rm c}$ due to the
behavior above $T_{\rm c}$ -- see below.)
The parameters in $B(t)$ then reflect {\it changes} in the magnetic
environment seen by the muons.

Results of fits are expressed as moments of the field distribution.
First, second, and third moments for the applied field of $100\,$Oe are
shown in Fig. \ref{fig2} as diamonds. 
\begin{figure}[htb]
\includegraphics[width=7.5cm]{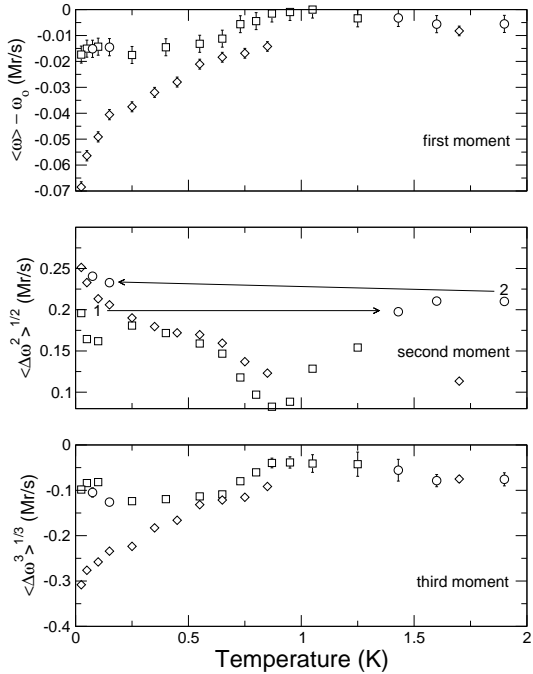}
\caption{Plot of the first, second, and third moments of
the field distributions derived from the back to back exponential
function fits: $100\,$Oe applied field (diamonds), $190\,$Oe applied field
(circles, squares).}
\label{fig2}
\end{figure}
The onset of superconductivity is at $T_{\rm c} \simeq 0.87\,$K,
a little lower than the {\it relaxed} state discussed above, and consistent
with
earlier studies of {\it intermediate} state samples\cite{Schwenk}.  One can see
immediately from the second moment data that the
behavior as $T \rightarrow 0\,$K is different than the {\it relaxed} sample
data
as well as what conventional BCS theory predicts.  The increasing second
moment denotes very unusual
behavior, which we attribute to the mixed phase state of the
sample.

The first and third moment graphs show increases in
diamagnetic shift and skewness, respectively, as $T\rightarrow 0\,$K.
Contrary to expectations, the skew is toward lower fields.  
This lower field skewness may be, for example, seen directly in
Fig. \ref{fig3}.  The tail to the left of the peak is responsible for the
skewness.
\begin{figure}[htb]
\includegraphics[width=7.5cm]{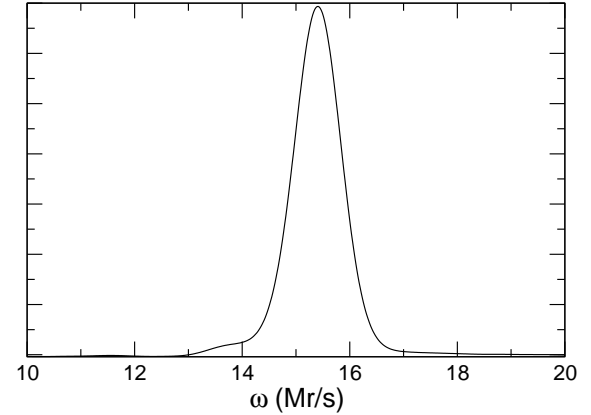}
\caption{The lineshape at $T=0.025\,$K for $190\,$Oe showing the low field
tail giving the negative third moment.}
\label{fig3}
\end{figure}
The lower field tail is not predicted for either a triangular or square
flux lattice alone\cite{Brandt}, and is most likely due to the SDW
phase.

The results of fits to data with $H_{ext}=190\,$Oe applied parallel to
the crystal {\bf c} axis are also shown in Fig. \ref{fig2}.
The second moment graph (circles and squares) again clearly shows the onset
of superconductivity at $T_{\rm c} \simeq 0.87\,$K.  Below this temperature
the field distribution broadens as fluxons in the superconducting phase form
into some type of lattice.  Above $T_{\rm c}$ the fits surprisingly show
evidence of a broad field distribution which increases with temperature.
This must be due to the SDW phase, since the superconducting regions are not
present here.  As $T_{\rm c}$ is approached from above, the field
distribution seen by the muons becomes more uniform, reflecting the decreasing
influence of the SDW phases in favor of the uniform, applied field.
Below $T_{\rm c}$ the behavior is generally the same as for the $100\,$Oe
data, but with more
scatter at low temperature.  The reason for this scatter, we think, is due to
the order in which the data were taken.  The square points indicate data taken
while lowering the temperature monotonically from $1.25\,$K to $0.025\,$K.
From $0.025\,$K the sample was warmed up to $1.43\,$K as indicated
by arrow 1.  The circles show results from fits as the sample was
warmed to $1.90\,$K.  Finally, the sample was cooled (more circles) to
$0.15$ and $0.075\,$K as indicated by arrow 2.  These last
points are higher than the previous low temperature data points for the
$190\,$Oe field, and not much higher than the circles above $T_{\rm c}$.
We believe this final higher second moment results due to the interplay
between the superconducting regions and the SDW regions.

The first and third moment graphs for $190\,$Oe show similar, but less
dramatic, effects than for $100\,$Oe.  In the first moment graph, the
diamagnetic shift is much smaller, which may indicate that a smaller
fraction of the sample is superconducting.  Similarly, the third moment
is less pronounced for this higher applied field.  Both of these effects
can be understood if one considers that the data fit by the first component
of the fitting function contain signals from
muons which stop in two different magnetic environments.  The amount of
signal from each is proportional to the number of muons stopping in
each, and a higher field should decrease the superconducting fraction
in favor of the SDW phase.

\section*{Conclusion}
We have studied the {\it relaxed} and some {\it intermediate} states of
the organic
superconductor $(TMTSF)_2ClO_4$ with $\mu$SR. While the error bars are too
large to rule out higher order quasiparticle pairing, the results of fits to
{\it relaxed} state data indicate a temperature dependence which appears to
be consistent with $s$-wave pairing.  The low temperature penetration depth
is found to be $\lambda_{ab}=12000\pm2000\,$\AA.
Recent, subsequent data taken on similar samples also reveal behavior
consistent with $s$-wave pairing, although the authors claim
otherwise\cite{Luke}.  The {\it intermediate} state, characterized by
coexisting regions of superconducting and SDW phases, has a suppressed
$T_{\rm c}$ and a lineshape with a low-field tail below $T_{\rm c}$.  Both of
these effects are proposed to be due to the existence of this SDW
phase in the material.

The authors would like to thank Mel Goode, Bassam Hitti, and the rest of the
TRIUMF support personnel for their help in making this work possible.


\begin{thebibliography}{9}
\bibitem{Belain}St\'{e}phane Belain and Kamran Behnis, Phys. Rev. Lett. 79 (1997) 2125.
\bibitem{Cooper}J.R. Cooper {\it et. al.}, Phys. Rev. Lett. 63 (1989) 1984.
\bibitem{Jerome}D. J\'{e}rome {\it et al.}, J. Phys. (Paris) Lett. 56 (1980) L-95.
\bibitem{Osada}T. Osada {\it et al.}, Phys. Rev. Lett. 66 (1991) 1525.
\bibitem{Hannahs}S.T. Hannahs {\it et al.}, Phys. Rev. Lett. 63 (1989) 1988.
\bibitem{Ishiguro}T. Ishiguro {\it et al.}, Organic Superconductors, (Spinger, Berlin, 1998).
\bibitem{Harsh1}D.R. Harshman and A.P. Mills, Phys. Rev. B45 (1992) 10684; D.R. Harshman, R.N. Kleiman, R.C. Hadden, W.J. Kossler, T. Pfiz, and D.Ll. Williams, unpublished; D.R. Harshman, Invited talk presented at the Gordon Research Conference on Organic Superconductivity, Irsee, Germany, 22-27 September, 1991.
\bibitem{Bechgaard}K. Bechgaard {\it et al.} Phys. Rev. Lett. 46 (1981) 852.
\bibitem{Schwenk}H. Schwenk {\it et al.}, Phys. Rev. B29 (1984) 500.
\bibitem{Garoche}P. Garoche {\it et al.}, J. Physique Lett. 43 (1982).
\bibitem{Chaikin}P.M. Chaikin {\it et al.}, J. Magn. Magn. Mater. 31 (1993) 1268.
\bibitem{Murata}K. Murata {\it et al.}, Japn. J. Appl. Phys. 26 (1987) 1367.
\bibitem{Choi}M.Y. Choi {\it et al.}, Phys. Rev. B 25 (1982) 6208.
\bibitem{Bouffard}S. Bouffard {\it et al.}, J. Phys. C: Solid State Phys. 15 (1982) 2951.
\bibitem{Coulon}C. Coulon {\it et al.}, J. Physique 43 (1982) 1721.
\bibitem{Tomic}S. Tomic {\it et al.}, J. Physique Coll. C3 (1983) 1075.
\bibitem{Lee2}I.J. Lee {\it et al.}, Synth. Metals 70 (1995) 747.
\bibitem{Lee3}I.J. Lee {\it et al.}, Phys. Rev. Lett. 78 (1997) 3555.
\bibitem{Lee4}I.J. Lee {\it et al.}, unpublished cond-mat/0001332.
\bibitem{Lee5}I.J. Lee {\it et al.}, Phys. Rev. Lett. 88 (2002) 17004.
\bibitem{Duncan}R.D. Duncan {\it et al.}, unpublished cond-mat/0102439.
\bibitem{Lebed}A.G. Lebed {\it et al.}, Phys. Rev. B62 (2000) R795.
\bibitem{Kuroki}K. Kuroki {\it et al.}, Phys. Rev. B63 (2001) 094509.
\bibitem{Miyazaki}M. Miyazaki {\it et al.}, unpublished cond-mat/9908488.
\bibitem{Schenck}A. Schenck, Muon Spin Rotation Spectroscopy: Principles and Applications in Solid State Physics (Hilger, Bristol, 1986).
\bibitem{Tinkham}M. Tinkham, Introduction to Superconductivity (McGraw-Hill, New York, 1975).
\bibitem{Brandt}E.H. Brandt, J. Low Temp. Phys. 73 (1988) 355.
\bibitem{Book}A.J. Greer and W.J. Kossler, Low magnetic fields in anisotropic superconductors (Springer-Verlag, Berlin, 1995).
\bibitem{Le}L.P. Le {\it et al.}, Phys. Rev. B48 (1993) 7284.
\bibitem{Harshman}D.R. Harshman {\it et al.}, Phys. Rev. Lett. 67 (1991) 3152.
\bibitem{Luke}G.M. Luke {\it et al.}, to be published in Physica B.
\end{thebibliography}
\end{document}